\documentstyle[aps,preprint,epsf]{revtex}

\begin{document}

\title{Tunneling between the Edges of Two Lateral Quantum Hall Systems}
\author{W. Kang$^{*,\S}$, H.L. Stormer$^{\dagger,\S}$, K.W. Baldwin$^{\S}$,
L.N. Pfeiffer$^{\S}$, and  K.W. West$^{\S}$}
\address{*James Franck Institute and Department of Physics,
 University of Chicago, Chicago, Illinois 60637\\}
\address{${\dagger}$Department of Physics, Columbia University, New York,
New York, 10027\\}
\address{${\S}$Bell Laboratories, Lucent Technologies, 600 Mountain
Avenue, Murray Hill, NJ 07974\\}

\maketitle
\narrowtext

\vspace*{1.5in}

{\bf The edge of a two-dimensional electron system (2DES) in a magnetic field 
consists of one-dimensional (1D) edge-channels that arise from the 
confining electric field at the edge of the specimen$^{1-3}$. 
The crossed electric and magnetic fields, E x B, cause electrons to drift parallel 
to the sample boundary creating a chiral current that travels along the edge 
in only one direction. Remarkably, in an ideal 2DES in the quantum Hall 
regime all current flows along the edge$^{4-6}$. Quantization of the 
Hall resistance, $R_{xy}= h/Ne^{2}$, arises from occupation of N 1D edge 
channels, each contributing a conductance of $e^{2}/h^{7-11}$. To explore 
this unusual one-dimensional property of an otherwise two-dimensional system, 
we have studied tunneling between the edges of 2DESs in the regime of 
integer quantum Hall effect (QHE). In the presence of an atomically precise, 
high-quality tunnel barrier, the resultant interaction between the edge 
states leads to the formation of new energy gaps and an intriguing dispersion 
relation for electrons traveling along the barrier. The absence of tunneling 
features due to the electron spin and the persistence of a conductance peak 
at zero bias are not consistent with a model of weakly interacting edge states. }

The edge channels of the QHE is a prototypical one-dimensional 
electronic system. Other such one-dimensional model systems include 
semiconductor based quantum wires$^{12}$, carbon nanotubes$^{13,14}$, and molecular 
chain materials$^{15}$.  While these systems force the electrons to move along 
a preferential direction, the edge channels of a QHE system is unique in 
its ability to adjust itself spatially as well as energetically. A localized 
defect is easily avoided by the edge current by simply skirting the 
impurity potential. In real samples, variations in the potential landscape 
can generate a complex and intriguing topology of edge 
channels$^{16,17}$. 
Consequently, most real edge channels are ill defined, both spatially and 
energetically, and it becomes difficult to generalize various experimental 
geometries in terms of 1D edge states. While typical experiments seek to study 
edge states in {\it lithographically defined} geometries$^{18-19}$, their edges 
typically fluctuate on the scale of a magnetic length. In order to address 
the energetics of edge channels, well-defined geometries with clearly 
delineated edges are essential. In particular for tunneling experiments, 
where distance factors exponentially into the tunneling current, exact 
knowledge of the shape and the magnitude of the barrier is crucial in making 
contact with model calculations.

    Our 2DES-barrier-2DES (2D-2D) tunneling device consists of two stripes 
of 2DESs separated by an atomically precise 88\AA-thick semiconductor barrier, 
fabricated by cleaved edge overgrowth, as shown in Fig.1a. We explore 
the lateral tunneling between these two physically separated 2DESs in 
the QHE regime. Fig.1c shows the differential conductance of the high density 
sample at 9.2T and 6.0T. At 9.2T, $dI/dV_{bias}$ is strongly suppressed around 
zero bias, while oscillatory conductance peaks appear above a threshold electric 
field. Each peak represents the onset of an additional tunneling path 
through the barrier. Successive conductance peaks are nearly equally spaced with
a spacing on the order of the cyclotron energy   $\hbar\omega_{c}$ ($\sim$17meV at 10T) 
and the amplitude of their oscillation decreases. Most of the $dI/dV_{bias}$-traces
resemble the 9.2T conductance data. However, for certain ranges of magnetic 
fields, there is no threshold to tunneling and a very sharp and tall conductance
peak arises at zero bias, as illustrated by the 6T data.  

\pagebreak

Fig.2 shows an image map of the $dI/dV_{bias}$ data as a function of Landau 
level filling factor, $\nu =nh/eB$, and the normalized bias voltage, 
$eV_{peak}/\hbar\omega_{c}$. Blue and red represent small and large 
$dI/dV_{bias}$ signals, respectively. With scaled axes, the disparate 
data from two different samples with very different electron densities join 
and produce a universal tunneling map for the 2D-2D tunneling. Altogether, 
an intriguing pattern arises with a continuous progression of conductance 
maxima (black squares and circles) from the low density to the high 
density data. The $dI/dV_{bias}$ plot exhibits gap-like thresholds 
that define regions of vanishing $dI/dV_{bias}$ in which the tunneling is 
strongly suppressed. Around zero bias, tunneling is suppressed in a bell-shaped 
region for filling factors $\nu < 1$. Similar thresholds occur at higher 
fillings near $\nu \sim 2$.  Between fillings $\nu = 1.2 - 
1.5$ and 2.2 - 2.5, a sharp conductance peak dominates at zero bias. 
The positions of the oscillatory conductance peaks vary roughly linearly
at high fillings and produce a fan of maxima that branches out from zero 
bias. Some of these branches cross each other above $\nu > 1$. For fillings $\nu < 1$ 
the different branches tend to saturate. 

    In analogy to the edge-state transport in the QHE, we consider the 2D-2D tunneling in 
terms of 1D edge states in the presence of a barrier. The edge states around the perimeter 
of the 2D-2D system are the same as in a simple 2DES, whose dispersion relation along 
the edge is approximately parabolic. This is no longer the case along the barrier. Fig. 3a 
illustrates the spatial dependence of the Landau level energy in its vicinity. The energy 
levels E(x) of Fig. 3a correspond to the energy of an electron with orbit guiding center $x$. 
Since $x = -k_{y}\ell_{\circ}^{2}$, the energy diagram of Fig. 3a is 
also the 1D {\it dispersion relation}, $E(k_{y})$, for 
electrons traveling {\it along} the barrier in the y-direction with velocity $v = h^{-1}dE/dk_{y} = 
h^{-1}\ell_{\circ}^{2}dE(x)/dx$.
 Electrons on the left side of the barrier counter-propagate with respect to those 
on the right side of the barrier. At the intersections of the two sets of rising Landau 
ladders, there exist two oppositely traveling, degenerate edge states with the same 
wavevector, $k_{y}= -x/\ell_{\circ}^{2}$. The degeneracy at the crossings is lifted by 
the formation of a 
series of small energy gaps that separate the symmetric and antisymmetric combinations 
of the underlying wavefunctions as seen in Fig. 3a. Altogether, Fig 3a bears a surprising 
resemblance with the data of Fig. 2.

Electronic transport {\it along} the barrier depends crucially on the position of the Fermi 
energy with respect to this complex 1D barrier level structure$^{20,21}$. 
When the Fermi level resides below the first maximum, say at $E_{1}$ 
in Fig. 3a, electrons in both 2DESs follow two 
separate, counter-propagating tracks along the junction as depicted in inset A of Fig. 3a. 
Although traveling along the barrier, they have very different $k_{y}$-wave vectors and are 
practically uncoupled. Therefore, tunneling through the barrier is negligible, corresponding 
to the absence of a peak at zero bias at low filling factor.

As the Fermi level reaches the first maximum at $E_{2}$, the edge states on both sides of the 
barrier have identical wavevector and resonate, which leads to substantial tunneling. In 
fact, the tracks of the wavefunctions have insignificantly changed as compared to inset A, 
but coupling between both edges has vastly increased due to the equivalence of their 
$k_{y}$-momentum. Consequently, at this particular position of  $E_{F}$, 
the conductance at zero bias 
turns finite, as observed in the zero bias peak of our data. According to the barrier level 
scheme in Fig. 3a, the edge states from the second Landau level, N = 1, are also occupied 
for this range of fillings, but remain uncoupled.

As the Fermi level rises slightly above this critical position and into the gap of  the 
dispersion relation, electrons can no longer travel {\it along} the barrier, as shown in the inset 
B of Fig. 3a. Consequently, coupling between both 2DESs should vanish and tunneling 
should cease. This represents an interesting paradox. If electrons can no longer travel 
along the barrier and, due to the chirality of the edge channel, are not allowed to 
backscatter, one would think they {\it need}  to tunnel. This should give rise to a conductance 
of $\sim e^{2}/h$, which is, however, 
two orders of magnitude larger than observed. The detailed 
tunneling and scattering model at this position of $E_{F}$ remains unclear. In any case, our 
model does not seem to agree with experiment, in which the zero-bias peak persists for a 
range of filling factors $1.2 < \nu <1.5$,  rather than existing at just one B-field. This may be 
related to electron scattering along the barrier, which relaxes k-conservation and washes 
out the dispersion relation. It is also unclear why the conductance vanishes abruptly at 
$\nu \sim 1.5$. This may coincide with $E_{F}$ reaching the top of the energy gap of the N = 0 branch, 
where both 2DESs couple once again along the barrier. A sharp zero-bias conduction and 
a strong suppression of tunneling alternate as $E_{F}$ moves through the higher lying gaps in 
Fig. 3a. 

   What is the origin of the complex overall pattern of Fig. 2, away from $V_{bias} = 0$, and what 
is its relationship to Fig. 3a? At finite bias, one set of Landau levels of Fig. 3a is raised 
with respect to the other by an energy $eV_{bias}$, as shown in Fig. 3b. This results in a shift 
of the intercepts and the conditions for onset of conduction is moved to a different 
energy. For example, when $E_{F} = E_{1}$ in Fig. 3a, tunneling across the barrier 
is inhibited, due 
to the absence of coupling of the edge states. In Fig. 3b the application of $eV_{bias}$ has 
shifted the crossing so as to coincide with $E_{1}$. This immensely enhances the coupling 
between both 2DESs in analogy to position $E_{2}$ in Fig. 3a and provides an explanation for 
the shift of the onset of tunneling to higher $V_{bias}$ for smaller $\nu$ in Fig. 2.

     In general, whenever the Fermi energy coincides with one of the crossing points, 
electrons can tunnel through the barrier and relax to the Fermi energy of the 2DES, lying 
$eV_{bias}$ below. Each additional coincidence of the levels produces a peak in $dI/dV_{bias}$. We 
can track their positions if we neglect the small gaps at the crossings. The Landau ladders 
on both sides are approximately described by
\begin{eqnarray}
\nonumber 
{E_{L} \over \hbar\omega_{c} } & = & \left( {x \over \ell_{\circ}}  + 
\sqrt{N_{L} + 1} \right)^{2} + (N_{L} + {1 \over 2})  \\
 &  &{\rm and } \\
\nonumber 
{E_{R} \over \hbar\omega_{c} } & = & \left( {x \over \ell_{\circ}}  -
\sqrt{N_{R} + 1} \right)^{2}¥ + (N_{R} + {1 \over 2}) - {eV_{bias} \over \hbar\omega_{c}}
\end{eqnarray}
continued by flat Landau levels for   ${x \over 
\ell_{\circ}} \geq \sqrt{N+1}$.
Neglecting $N_{R}$ and $N_{L}$ versus $eV_{bias}/2\hbar\omega_{c}$, 
which is justified for large sections of Fig. 2, and assuming that
${E_{L} \over \hbar\omega_{c} } \approx \nu$, which holds for very broad 
Landau levels, as is likely  the case in our lower-mobility specimen, 
we arrive at 
\begin{eqnarray}
\nu \approx \left( {eV_{bias} \over 2\hbar\omega_{c}}  + 
\sqrt{N_{L} + 1} \right)^{2} + (N_{L} + {1 \over 2}).
\end{eqnarray}
as the condition for coincidence. Eq. 1, which, together with 
its $L-R$ mirror image describes the barrier level structure of 
Fig. 3a, is identical to eq. 2, which identifies the peaks in 
the differential conductance in Fig. 2 as long as 
${E_{L} \over \hbar\omega_{c}}$  is replaced by $\nu$ and $x/\ell_{\circ}$
is replaced by $eV_{bias}/2\hbar\omega_{c}$. 
This 
resolves the puzzle of the intriguing similarity between both graphs.

     While we can account for the general features of our experimental results, many 
aspects of the data remain unresolved and require an explanation beyond our simple 
model. One such feature is the extended existence of the zero-bias peak, which is expected 
to occur only at the point of coincidence of $E_{F}$ with the edges of the gaps. The origin of 
this discrepancy may be a relaxed $k$-conservation due to remnant disorder along the 
barrier, which scatters electrons, broadens the dispersion relation and therefore extends 
the allowed energy range for strong tunneling. Such a broadening may also account for the 
conductance which is much reduced compared to $e^{2}/h$: The sharp resonance with universal 
conductance is broadened into a band of much lower, average conductance. However, the 
sharpness of the zero-bias peak as a function of voltage bias implies limited broadening. A 
proper accounting of the role of disorder in 2D-2D tunneling requires a detailed, 
quantitative analysis of our 2D-2D device.

Another puzzling feature of our data is the position in filling factor at which the zero 
bias conductance peak appears.  According to Fig. 3a, the first coincidence of the Landau 
ladders occurs at  $\nu > 4$, somewhat above the N = 1 Landau level, contrary to the 
experimental value of $\nu \approx 1.2$. 
Similarly, the next coincidence is expected for $\nu >6$, while it 
is observed at $\nu \sim 2.2$. 
These observations point to a shifting of the levels in addition to a 
possible broadening. It could arise from self-consistent screening and from an 
accumulation of charge in the vicinity of the barrier$^{22}$, which modifies the level scheme. 

   Finally, the influence of the electron spin on our experiment as well as on the dispersion 
relation in Fig. 3a remains unclear.  While the Zeeman splitting in GaAs is only ~1/70 of 
the cyclotron splitting, partial and spatially dependent occupation of Landau levels near 
the barrier can produce large spatially dependent enhancement of the 
$g$-factor$^{1,2}$. This can 
give rise to additional coincidences -- possibly over ranges of fillings Ð and will appear in 
the tunneling characteristics, in particular, if spin-flip scattering is strong. This absence of 
spin-dependent features in the tunneling data remains a key puzzle. Numerical studies of 
our comparably simple physical system should provide guidance as to the relative 
importance of different mechanisms.

\vspace*{0.25in}

\noindent
---------------------------\\
1. Prange, R.E. \& Girvin, S.M. (eds) {\it The Quantum Hall Effect} 2nd edn (Springer, New 
York, 1990).\\
2. Das Sarma, S. \& Pinczuk, A. (eds) {\it Perspectives in Quantum Hall Effects} (Wiley Inter-
Science, New York, 1997).\\
3. Halperin, B.I. Quantized Hall conductance, current-carrying edge states,
and the existence of extended states in a two-dimensional disordered potential.
Phys. Rev. B  {\bf 25}, 2185-2188 (1983).\\
4.MacDonald, A.H. \& Streda, P. Quantized Hall effect and edge currents. 
Phys. Rev. B {\bf 29}, 1616-1619 (1984).\\
5. Apenko, S.M. \& Lozovik, Yu. E. J. The quantized Hall effect in strong magnetic fields. 
Phys. C {\bf 18}, 1197-1203 (1985).\\
6. Fontein, P.F. et al. Spatial potential distribution in 
GaAs/Al$_{x}$Ga$_{1-x}$As heterostructures 
under quantum Hall conditions studied with the linear electro-optic effect. Phys. Rev.
B {\bf 43}, 12090-3 (1991).\\
7. Buttiker, M.  Absence of backscattering in the quantum Hall 
effect in multiprobe conductors. Phys. Rev. B {\bf 38}, 9375-9389 
(1988).\\
8. Streda, P., Kucera, J. \& MacDonald, A.H. Edge states, transmission matrices, and the 
Hall resistance. Phys. Rev. Lett. {\bf 59}, 1973-1975 (1987).\\
9. Jain, J.K. \& Kivelson, S.A. Landauer-type formulation of quantum-Hall transport: 
critical currents and narrow channels. Phys. Rev. B {\bf 37}, 4276-4279 
(1988).\\
10. Haug, R.J., MacDonald, A.H., Streda, P. \& von Klitzing. Quantized multichannel 
magnetotransport through a barrier in two dimensions. K. Phys. Rev. Lett.{\bf  61}, 2797-2800 
(1988).\\
11. Washburn, S., Fowler, A.B., Schmid, H. \& Kern, D. Quantized Hall effect in the 
presence of backscattering. Phys. Rev. Lett. {\bf  61}, 2801-2804 
(1988).\\
12. Yacoby, A. et. al. Non-universal conductance quantization in quantum wires. Phys. 
Rev. Lett. {\bf 77}, 4612-4615 (1996).\\
13. Wildoer, J.W. G., Venema, L.C., Rinzler, A.G., Smalley, \& R.E., Dekker, C. 
Electronic structure of atomically resolved carbon nanotubes. Nature {\bf 
391}, 59-62 (1998).\\
14. Odom, T.W., Huang, J., Kim, P. \& Lieber, C.M. Atomic structure and electronic 
properties of single walled carbon nanotubes. Nature {\bf 391}, 62-64 
(1998).\\
15. Ishiguro, T., Yamaji, K., \& Saito, G. {\it Organic Supercondcutors} 2nd edn  (Springer-
Verlag, New York, 1998). \\
16. Tessmer, S.H., Glicofridis, P.I., Ashoori, R.C., Levitov, L.S., \& Melloch, M.R. 
Surface charge accumulation imaging of a quantum Hall liquids. Nature. 
{\bf 392}, 51-54 (1998).\\
17. McCormick, K.L. et. al. Scanned potential microscopy of edge and bulk currents in 
the quantum Hall regime. Phys. Rev. B {\bf 59}, 4654 (1999).\\
18. Goldman, V.J. \& Su, B. Resonant tunneling in the quantum Hall regime: measurement 
of fractional charge. Science {\bf 267}, 1010-12 (1995).\\
19. Tarucha, S., Honda, T., \& Saku, T. Reduction of quantized conductance at low 
temperatures observed in 2 to 10 $\mu m$-long quantum wires. Solid State Communications, 
{\bf 94}, 413-18 (1995).\\
20. Ho, T.L. Oscillatory tunneling between quantum Hall systems. Phys. Rev. B {\bf 50}, 
4524-4533 (1994).  \\
21. Girvin, S.M. Private Communication.\\
22. Wulf, U., Gudmundsson, V. \& Gerhardts, R.R. Screening properties of the two-
dimensional electron gas in the quantum Hall regime. Phys. Rev. B. {\bf 
38}, 4218-4230 (1988).\\
23. Pfeiffer, L.N. et. al. Formation of a high quality two-dimensional electron gas on 
cleaved GaAs. Appl. Phys. Lett  {\bf 56}, 1697-1699 (1990).\\
24. Chang, A.M., Pfeiffer, L.N., \& West, K.W. Observation of chiral Luttinger behavior 
in electron tunneling into fractional quantum Hall edges. Phys. Rev. Lett. {\bf 77}, 2538-2341 
(1996).\\

\pagebreak

\noindent
\hrulefill\\
We are very grateful to S.M. Girvin for providing us with much insight 
 into the intricate energetics of our experimental geometry. We would also 
 like to thank R. De Picciotto, A. M.  Chang, T.L. Ho, and J.P. Eisenstein 
 for valuable discussions.

\vspace*{0.1in}

\noindent
Correspondence and requests for materials should be addressed to 
W.K.(e-mail: wkang@rainbow.uchicago.edu).

\vspace*{0.2in}

\noindent
Fig. 1.  Structure and differential conductance measurement of the 2D-2D tunneling device. 
(a). The junctions are fabricated by cleaved edge overgrowth in molecular beam epitaxy 
(MBE)$^{23,24}$. The first growth on a standard (100) GaAs substrate consists of an undoped 
13$\mu m$ GaAs layer followed by a  88\AA-thick digital alloy of undoped Al$_{0.1}$Ga$_{0.9}$As/AlAs,
and completed by a 14$\mu m$ layer of undoped GaAs. This triple-layer sample is cleaved 
along the (110) plane in an MBE machine and a modulation-doping sequence is grown 
over the exposed edge. It consists of a 3500\AA-thick AlGaAs layer, delta-doped with Si at 
a distance of 500\AA\ from the interface. Carriers from the Si impurities transfer only into 
the GaAs layers of the cleaved edge, forming two stripes of 2DESs of width 13$\mu m$ and 
14$\mu m$ separated from each other by a 88A-thick, Al$_{0.1}$Ga$_{0.9}$As/AlAs barrier. 
The sample is fabricated into a mesa incorporating the barrier and the two 2DESs. 
Contacts are made to the 2DESs, far away from the tunneling region. (b) Schematic band 
structure of the 2D-2D tunneling device. Two different samples with electron density of 
$n = 1.1\times 10^{11} cm^{-2}$ and 
$n = 2.0\times 10^{11}cm^{-2}$ are studied. From a simultaneously grown 
monitor wafer we estimate the mobility of the 2DESs in the device to be 
$\sim 1\times 10^{5} cm^{2}/Vsec$.
(c) Two representative traces of differential conductance through the tunneling 
barrier with electron density $n = 2.0\times 10^{11}cm^{-2}$. 
A low-frequency AC-technique (typically 
10 $\mu V$ amplitude) is employed to measure the differential conductance, 
$dI/dV_{bias}$, through 
the barrier in the presence of a DC voltage bias, $V_{bias}$. The samples are measured at T = 
300mK in a magnitude field. Near zero bias differential conductances, $dI/dV_{bias}$, nearly 
vanish, while conductance oscillations on the order of 
$10^{-6}\Omega^{-1}$ are found at high bias. 

\pagebreak

\noindent
Fig. 2. 
 Evolution of differential conductance of 2D-2D tunneling devices under high 
magnetic fields. The Landau level filling factor, $\nu =nh/eB$, and the normalized bias,  
$eV_{peak}/\hbar\omega_{c}$, are used as universal axis for the 2D-2D tunneling data 
from samples with 
electron densities, $n = 2.0\times 10^{11}cm^{-2}$  (top) and 
$n = 1.1\times 10^{11}cm^{-2}$ (bottom). Their maxima are 
indicated by black circles and squares, respectively. Blue (red) regions represent minimum 
(maximum) conductance. In the blue region, conductances,  $dI/dV_{bias}$,nearly vanish, while 
the red areas represent conductances on the order of $10^{-6}\Omega^{-1}$.  

\vspace*{0.20in}
\noindent
Fig. 3. Schematic energy dependence of the Landau levels in the vicinity of the barrier. (a). 
Shown for zero voltage bias. Far away from the barrier Landau levels are equally spaced, 
$E_{N} = (N+1/2)\hbar\omega_{c}$. As electrons approach the edge or the barrier their energy rises 
parabolically.
\begin{eqnarray}
\nonumber
{E \over \hbar\omega_{c} } = \left( \left| {x \over \ell_{\circ}} \right| + 
\sqrt{N + 1} \right)^{2} + (N + {1 \over 2}), N = 0,1,2,\ldots 
{\rm and} \left| {x \over \ell_{\circ}}\right| \leq \sqrt{N + 1},
\end{eqnarray}
{\rm otherwise}

${E \over \hbar\omega_{c} } = N + {1 \over 2} $
where $\ell_{\circ} = \sqrt{eB/h}$  is the magnetic length. 
In the vicinity of the barrier, these parabolas anti-cross and create a 
gapped spectrum. The traces represent the energy of an electron whose
guiding center is located at $x$. N is the Landau level index, $E_{1}$ 
through $E_{3}$ represent Fermi energies at different filling factors. 
Inserts represent the in-plane track of the electronic wavefunction. 
(b) Same as Fig. 3a but with a bias of $\sim 2\hbar\omega_{c}$ 
applied across the barrier.

\end{document}